\documentclass[11pt,titlepage]{article}
\usepackage{a4wide,epsfig,amssymb}

\newcounter{nref}
\newcommand{\bbib}{%
  \renewcommand{\refname}{\large\bf References}%
  \begin{thebibliography}{99}%
  \setcounter{enumiv}{\thenref}}
\newcommand{\ebib}{%
  \end{thebibliography}%
  \setcounter{nref}{\arabic{enumiv}}}
\newcommand{\head}[3]{%
  \setcounter{nref}{0}%
  \thispagestyle{empty}%
  \section*{\LARGE\bf #1}%
  \stepcounter{section}%
  \addcontentsline{toc}{section}{#1}%
  \large\itshape%
  #2\\\vspace{0.1pt}\\%
  #3%
  \normalsize\upshape%
  \bigskip}

\begin{document}


\head{Core-collapse supernova simulations:\\
      Variations of the input physics}
     {M.\ Rampp$^1$, R.\ Buras$^1$, H.-Th.\ Janka$^1$, G.\ Raffelt$^2$}
     {$^1$ Max-Planck-Institut f\"ur Astrophysik,
     Karl-Schwarzschild-Str.~1, 
     D-85741 Garching, Germany\\
      $^2$ Max-Planck-Institut f\"ur Physik, F\"ohringer Ring 6, 
     D-80805 M\"unchen, Germany}

\subsection*{Abstract}

Spherically symmetric simulations of stellar core collapse and
post-bounce evolution are used to test the sensitivity of the
supernova dynamics to different variations of the input physics.
We consider a state-of-the-art description of the neutrino-nucleon
interactions, possible lepton-number changing neutrino reactions in
the neutron star, and the potential impact of hydrodynamic mixing
behind the supernova shock.

\subsection*{Improvements of the neutrino-nucleon interaction rates}

Recently, spherically symmetric Newtonian \cite{rampp.ramjan00,rampp.mezlie01},
and general relativistic \cite{rampp.liemez01} hydrodynamical
simulations of stellar core-collapse and the post-bounce evolution
including a Boltzmann solver for the neutrino transport
have become possible. No supernova explosions were obtained.

Having reached a new level of accuracy of the numerical treatment,
it is a natural next step to reconsider the input physics that enters the
models, remove imponderabilities of their description and also test
possible alternatives.
Within the set of so-called ``standard''  neutrino opacities
\cite{rampp.bru85} which have been widely used in supernova models so far, for example, 
the rates of charged-current
interactions of electron  neutrinos and antineutrinos with free
nucleons as well as
neutral-current scatterings of neutrinos off free nucleons were
calculated with the assumption that the
nucleons can be considered as isolated, infinitely massive particles at rest.
It is, however, well known that these interaction rates are
significantly changed when energy transfer between the nucleons and
leptons (``recoil''), nucleon-nucleon
correlations due to Fermi statistics and nuclear forces, and other so
far disregarded effects like  
weak magnetism corrections, are adequately taken into account (see
Ref.~\cite{rampp.hor02}). 
Effectively, at densities $\rho\gtrsim 10^{13}\,\mathrm{g}/\mathrm{cm}^3$,
the opacities become considerably smaller than in the
``standard'' approximation.  
Correspondingly, neutrino diffusion through the nascent neutron
star and the emission from  the neutrinosphere are 
enhanced and higher neutrino luminosities must be expected.

\begin{figure}[ht]
   \centerline{
     \put(0.9,0.3){{\Large\bf a}}
     \epsfclipon\epsfxsize=0.45\textwidth\epsffile{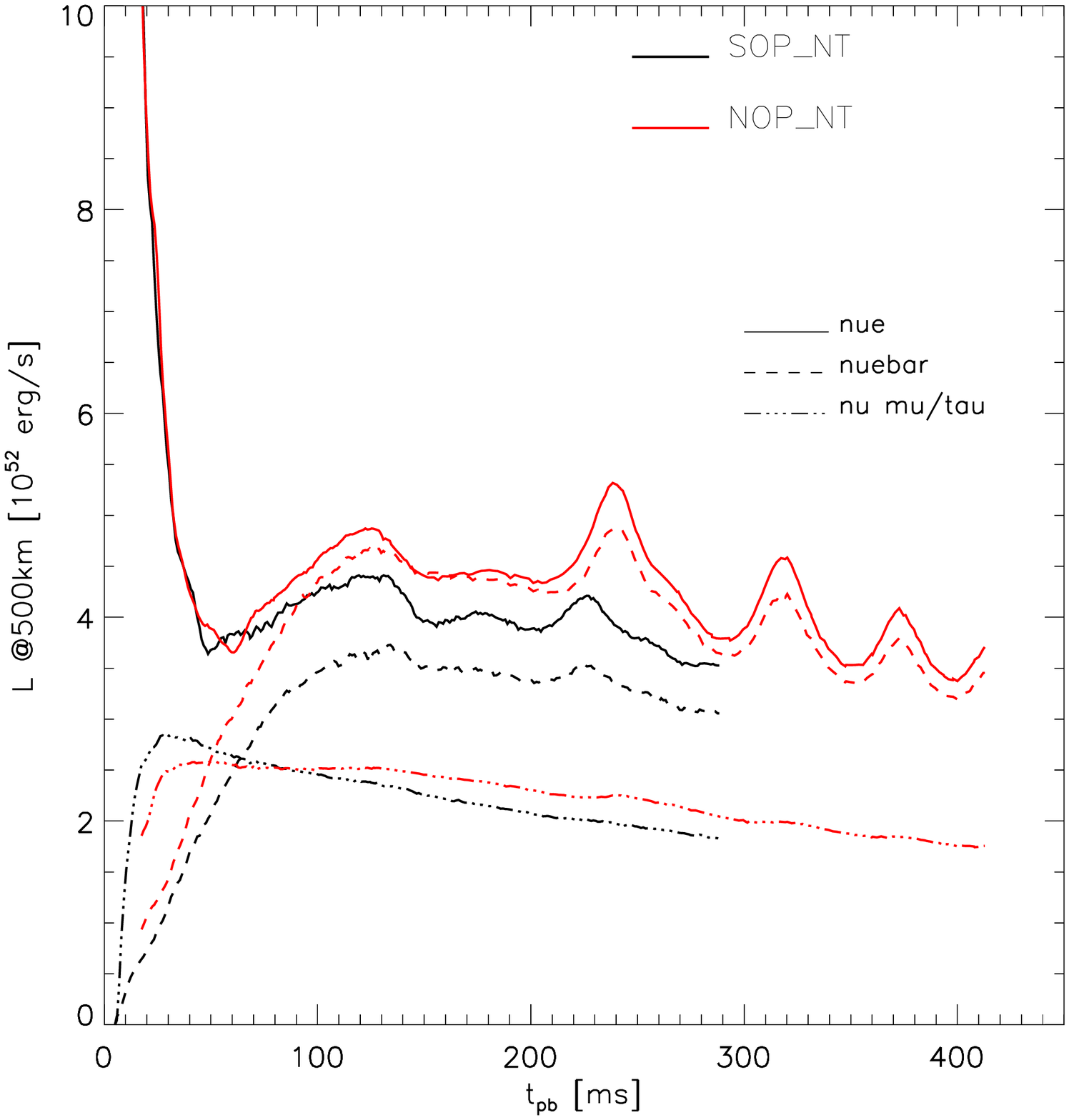}
     \put(0.9,0.3){{\Large\bf b}}
     \epsfclipon\epsfxsize=0.45\textwidth\epsffile{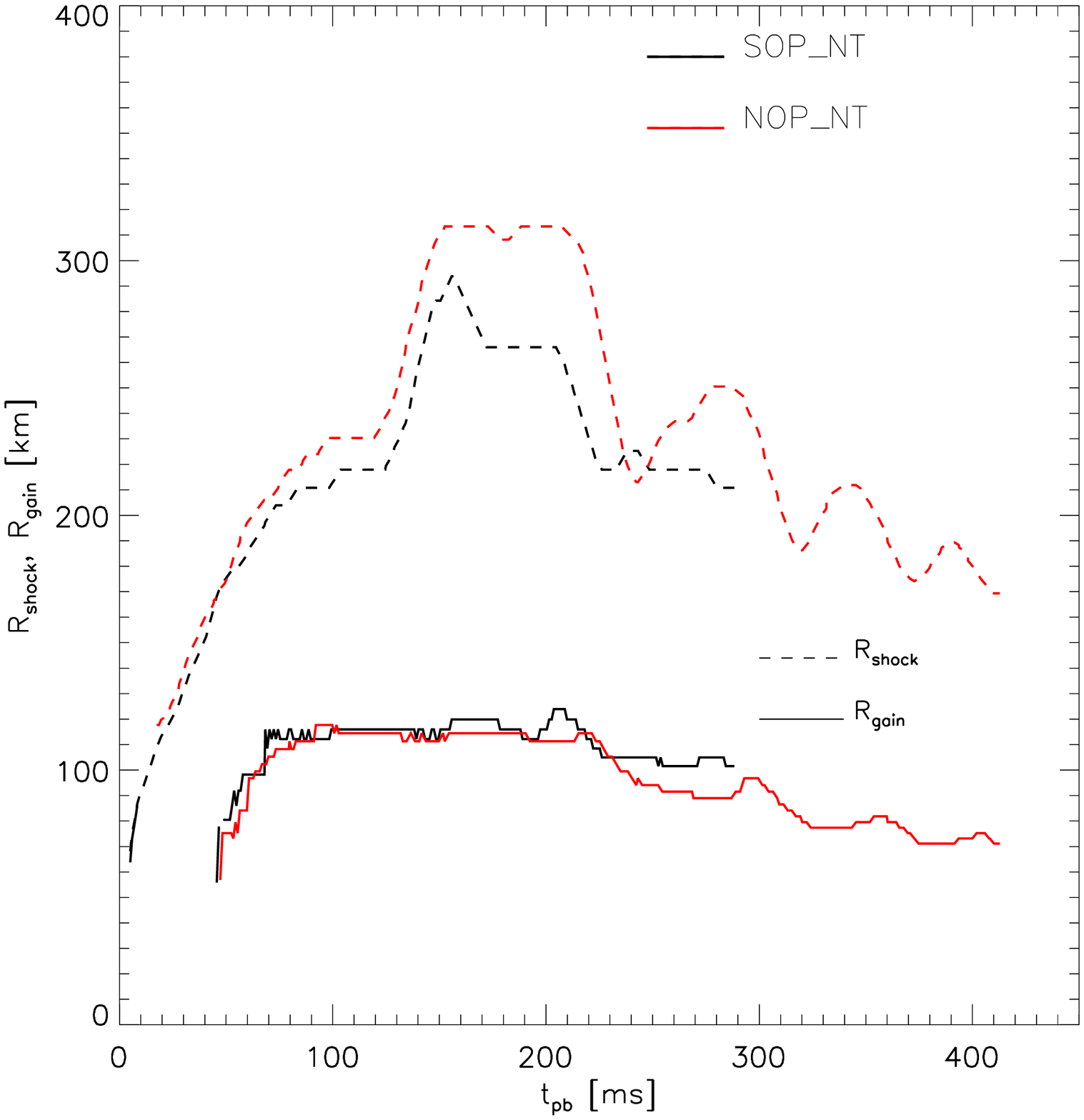}
     }
  \caption{Comparison of the post-bounce evolution of a model computed
    with ``standard'' neutrino opacities (black lines) and
    a model which employs the improved description of neutrino-nucleon
    interactions (red lines).
    Panel {\bf a} shows the luminosities of $\nu_e$ (solid lines), 
    $\bar\nu_e$ (dashed lines), and $\nu_\mu$, $\nu_\tau$, $\bar\nu_\mu$,
    $\bar\nu_\tau$ (individually; dash-dotted lines) as functions of
    time. In Panel {\bf b} the radial positions of the shock
    (dashed lines) and the gain radius (solid lines) of the
    two models are compared.}  
  \label{rampp.fig1}
\end{figure}

Figure~\ref{rampp.fig1} shows results of Newtonian simulations
of iron core-collapse and the post-bounce evolution  of 
a $15\,M_{\odot}$ star
\cite{rampp.woo99} using Boltzmann neutrino transport \cite{rampp.ramjan02}.  
One model was calculated with an improved description of
neutrino-nucleon interactions, which includes  the detailed reaction
kinematics and nucleon phase-space blocking, nucleon-nucleon
correlations in the dense medium 
\cite{rampp.redpra98,rampp.bursaw98}, weak magnetism \cite{rampp.hor02} and  the
possible quenching of the axial coupling $g_A$ in nuclear matter
\cite{rampp.carpra02}.  
For a reference calculation the conventional ``standard''
opacities \cite{rampp.bru85} were used. In both simulations we also
included  nucleon-nucleon bremsstrahlung $NN \rightleftharpoons
\nu\bar\nu\,NN$ ($N\in\{n,p\}$ denotes a free nucleon),
which is the dominant production reaction of $\mu$ and $\tau$
neutrinos and antineutrinos in the denser regions of the newly formed
neutron star \cite{rampp.hanraf98,rampp.thobur00}.  

As expected from the arguments above, the model calculated with the
new implementation of the opacities shows enhanced
luminosities 
for $\nu_e$ and, even more pronounced, for $\bar\nu_e$. 
This is caused by the fact that for $\nu_e$ and $\bar\nu_e$ recoil and
nucleon correlations have the strongest effect at high densities,
while for $\bar\nu_e$ the weak magnetism reduces the opacities also at 
moderate densities 
$\rho\lesssim 10^{13}\,\mathrm{g}/\mathrm{cm}^3$ \cite{rampp.hor02},
where most of the neutrino emission is produced during the time of 
consideration. 
In contrast, for $\nu_e$ both ``corrections'' counteract and almost cancel
each other \cite{rampp.hor02}.
The higher luminosities (see Fig.~\ref{rampp.fig1}a)  and somewhat larger mean
energies of $\nu_e$ and $\bar\nu_e$ increase the neutrino heating in
the gain region below the hydrodynamic shock. 
Consequently, the shock propagates to larger radii when compared with
the reference model (see Fig.~\ref{rampp.fig1}b).

In view of the appreciable differences between both models we
consider the inclusion of a state-of-the-art description of
neutrino-nucleon interactions as a mandatory step
towards more realistic supernova simulations, although the improved opacities
\emph{alone} do not lead to a successful supernova explosion in spherical symmetry. 

\subsection*{Lepton-number changing neutrino reactions}

\begin{figure}[ht]
   \centerline{
     \put(0.9,0.3){{\Large\bf a}}
     \epsfclipon\epsfxsize=0.45\textwidth\epsffile{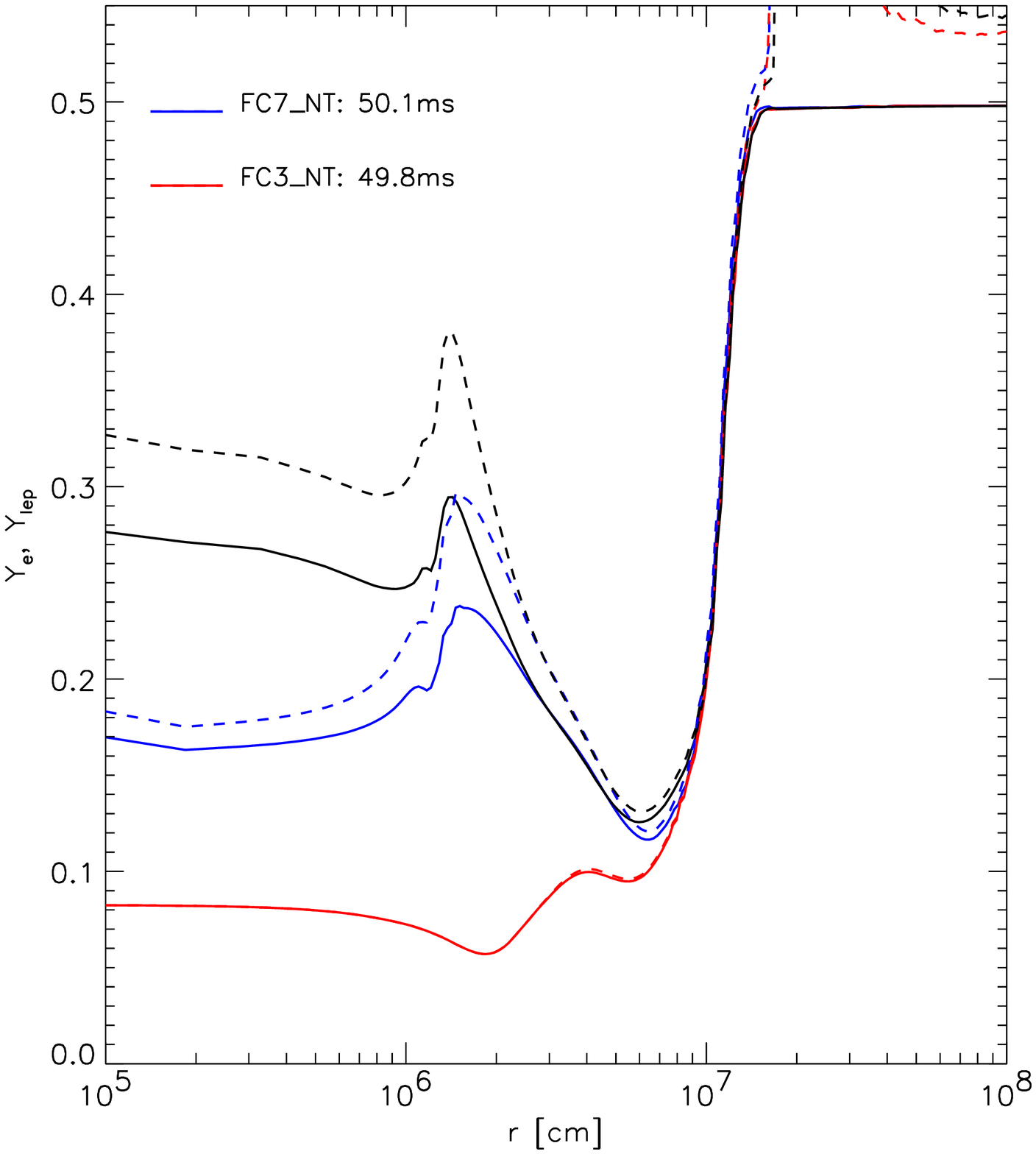}
     \put(0.9,0.3){{\Large\bf b}}
     \epsfclipon\epsfxsize=0.45\textwidth\epsffile{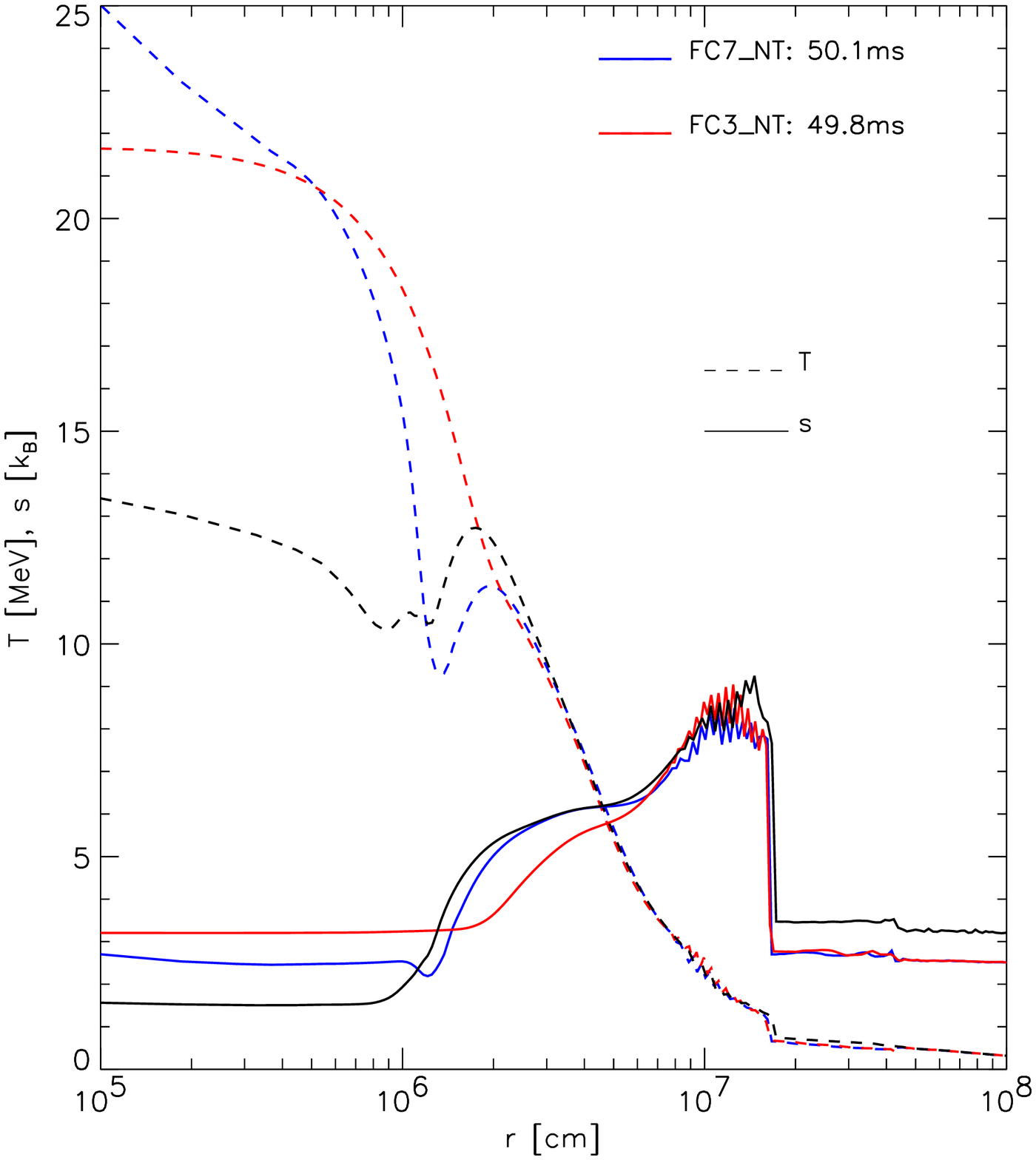}
     }
  \caption{Radial profiles of the electron fraction (Panel {\bf a},
    solid lines), total lepton fraction (Panel {\bf a}, dashed lines),
    temperature (Panel {\bf b}, dashed lines) and 
    specific entropy (Panel {\bf b}, solid lines) 50~ms after core bounce. 
    Results of the models with lepton-number changing
    reactions are plotted as 
    red ($\sigma/\sigma_{\mathrm{SM}}=10^{-3}$) and 
    blue lines ($\sigma/\sigma_{\mathrm{SM}}=10^{-7}$). 
    For comparison quantities of the reference model
    ($\sigma/\sigma_{\mathrm{SM}}=0$) are shown as black lines.}\label{rampp.fig2} 
\end{figure}

Lepton-number changing interactions are expected in various extensions
of the particle-physics standard model, notably in R-parity violating
models of supersymmetry; see Ref.~\cite{rampp.berkla00} for an
overview and current experimental limits.  Such processes might
significantly affect the supernova dynamics~\cite{rampp.fuller88}.  
As a proxy
for this class of interactions we implemented the reaction $\nu_e +N
\rightleftharpoons \bar\nu_e+N$ which opens a channel for ``internal
deleptonization'', i.e.\ quickly running down electron lepton number
without diffusion to the stellar surface.  This effect leads to a
corresponding increase of the temperature of the stellar plasma by
conversion of degeneracy energy to thermal energy
(Fig.~\ref{rampp.fig2}).  The relevant time scale is fixed by the
effective strength of the lepton-number changing reaction.  We
parametrize it by $\sigma/\sigma_{\mathrm{SM}}$, which is the
lepton-number changing cross section normalized to the standard-model
one for $\nu_e +N \rightleftharpoons \nu_e+N$.  We have calculated one
model with $\sigma/\sigma_{\mathrm{SM}}=10^{-7}$ where the
deleptonization time scale is too slow to be effective during infall,
and an extreme case where $\sigma/\sigma_{\mathrm{SM}}=10^{-3}$ and
where coherent enhancement by the scattering off heavy nuclei is
allowed, $\nu_e +A \rightleftharpoons \bar\nu_e +A$.
Despite of sizeable differences in the physical conditions in the core
of the nascent neutron star (Fig.~\ref{rampp.fig2}) we find a
remarkable insensitivity of the overall dynamical evolution of the
models to the dramatic modifications of the microphysics.  In both
models the hydrodynamic shock reaches a maximum radius of about 300~km
at $\approx 200$~ms after bounce. These values are very close to the
reference calculation with $\sigma/\sigma_{\mathrm{SM}}=0$.
This finding 
is probably explained by 
the accelerated  deleptonization and heating being confined
to the innermost 
$\approx 20$~km of the neutron star (Fig.~\ref{rampp.fig2}),
because only there the lepton-number changing reactions are fast
enough and the optical depth for these processes is larger than unity.
During the first few hundred milliseconds after bounce most of the
neutrino luminosity, on the other hand, originates from regions beween
$50$~km and $100$~km where the temperature is essentially the same in
all models (Fig.~\ref{rampp.fig2}b).  Continuing the simulations over
a period of nearly one second, we find slightly enhanced neutrino
luminosities and mean energies and thus heating in the gain region
at later times after bounce, but not to
a degree that the conditions for explosions~\cite{rampp.jan01} would
become more favourable.

\begin{figure}[ht]
   \centerline{
     \put(0.9,0.3){{\Large\bf a}}
     \epsfclipon\epsfxsize=0.45\textwidth\epsffile{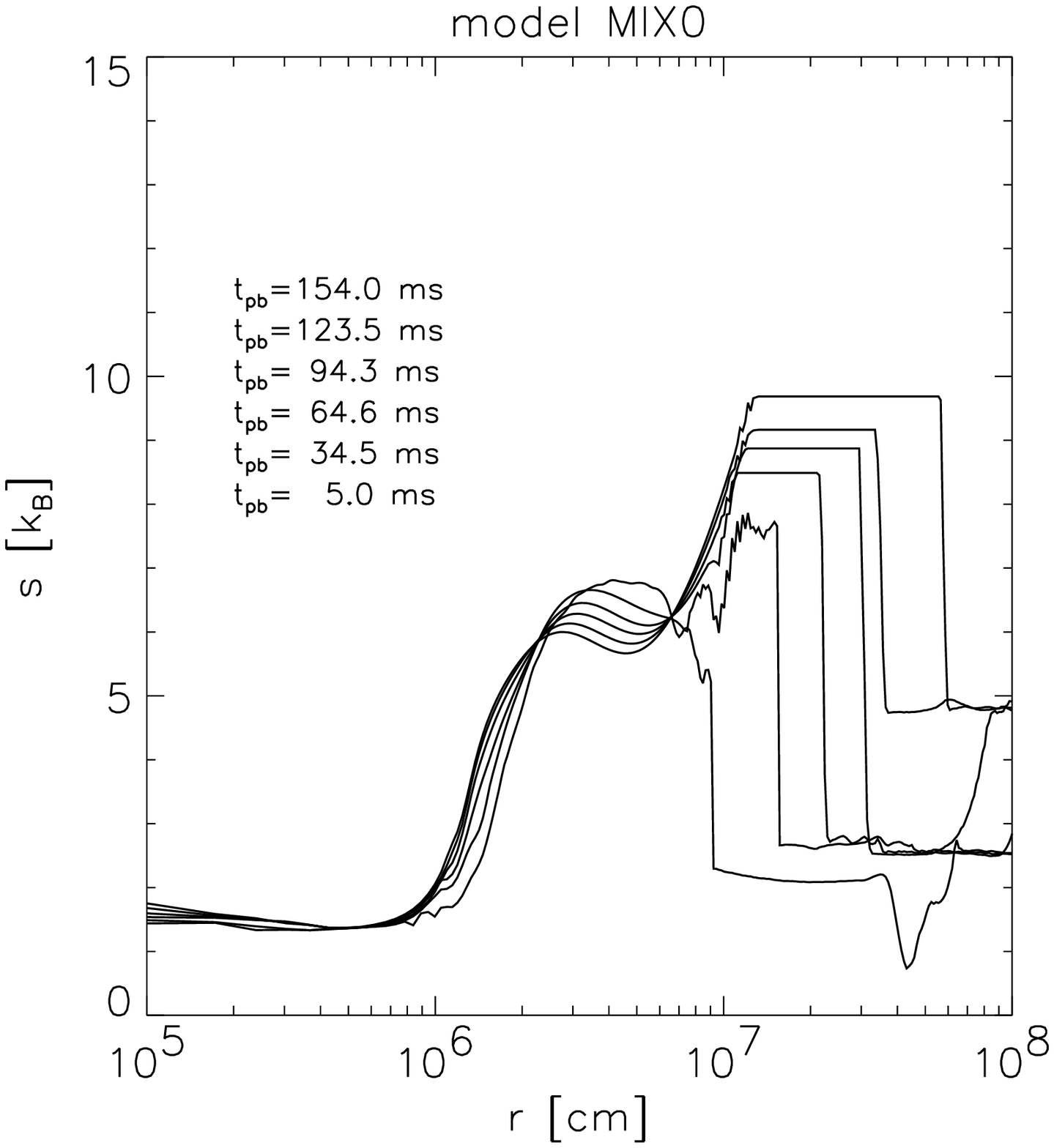}
     \put(0.9,0.3){{\Large\bf b}}
     \epsfclipon\epsfxsize=0.45\textwidth\epsffile{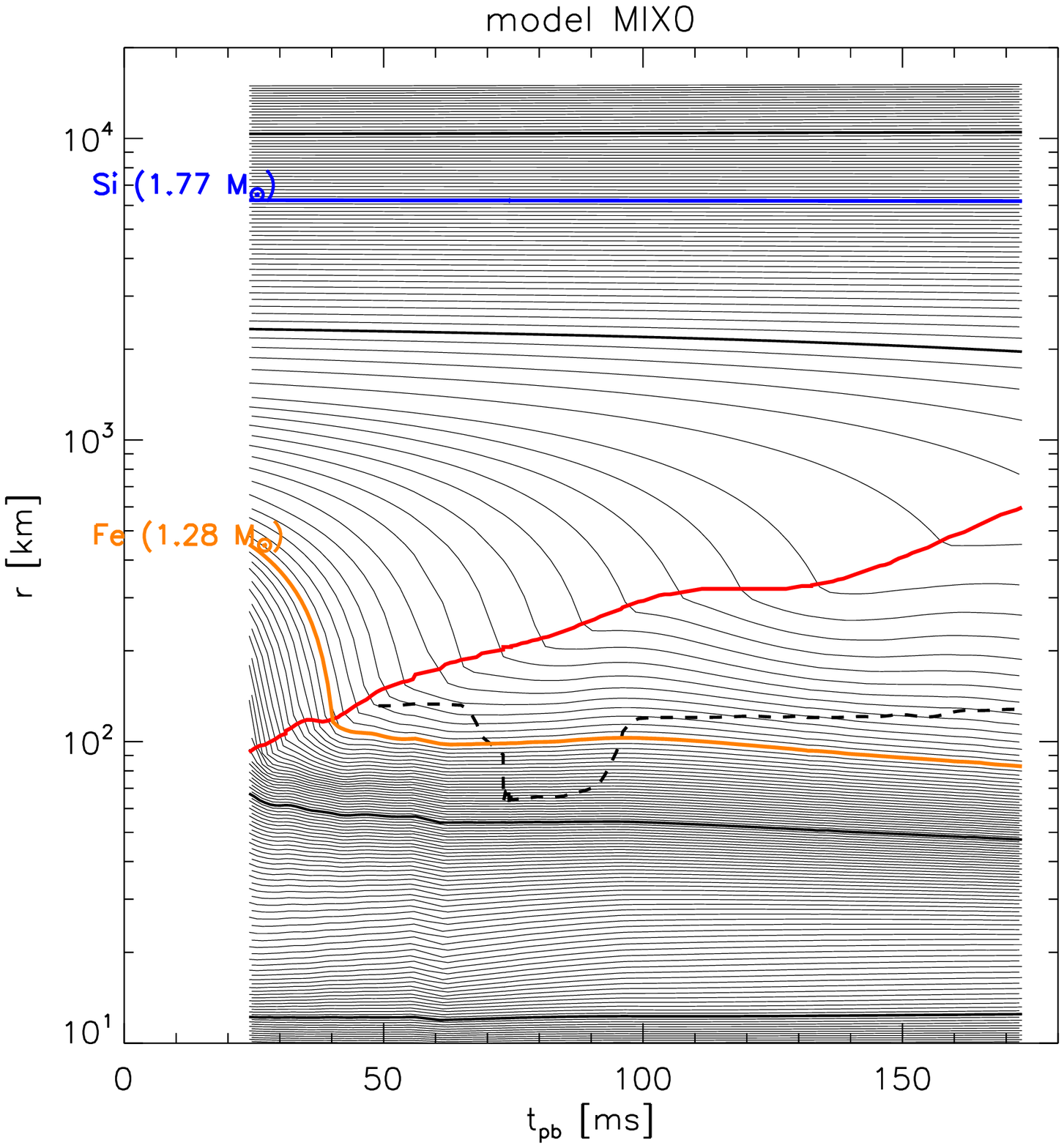}
     }
  \caption{Panel {\bf a} shows the effect of our ``mixing'' algorithm
    on the temporal evolution of the radial
    profile of the specific entropy. 
    From bottom to top the times correspond to
    profiles with increasing values of the postshock entropy.
    For the same model the trajectories of different
    mass shells (with a spacing of $0.02~M_\odot$) are plotted as
    black lines in Panel {\bf b}. 
    The red line traces the radial position of the supernova shock.
    Note that the model includes only $\nu_e$ and $\bar\nu_e$, and can
    be directly compared with the simulation published in
    Ref.~\cite{rampp.ramjan00}.}  
  \label{rampp.fig3}
\end{figure}

\subsection*{Successful explosions by convective processes ?}

Multi-dimensional hydrodynamic simulations, coupled with simplified 
treatments of the neutrino physics, have shown that large-scale
convective overturn behind the supernova shock \emph{can} aid the
neutrino-driven mechanism and 
cause an explosion even if models fail in spherical symmetry
(see Ref.~\cite{rampp.jankif01} for an overview and references 
to original work).
However, because of the uncertainties and approximations in the
description of 
the neutrino sector, the actual role of convection in supernova models
is not yet clear.
Self-consistent and multi-dimensional simulations with a sufficiently
accurate treatment of the neutrino transport have still to be
carried out.

Convective overturn has the helpful influence of suppressing energy
losses by the reemission of neutrinos, because neutrino-heated matter
expands outward instead of being advected through the gain region into
the cooling layer. In addition, part of the matter falling through the
shock can still accrete onto the neutron star, which helps keeping up a
high value of the accretion luminosity. Furthermore, the expanding,
neutrino-heated matter raises the pressure immediately behind the shock
and thus drives the shock farther out.
We try to mimick the latter effect of a multi-dimensional fluid flow
between the shock and the gain radius by
artificially (and instantaneously) mixing layers of negative entropy
gradient behind the shock in our spherically symmetric
simulations. This leads to a flattening of the entropy profile. 
We achieve this behaviour by setting the entropy in such a region to a
constant value $s(t,r)=s_0(t)$ (see Fig.~\ref{rampp.fig3}a), which is calculated 
in each timestep from the requirement that the internal (and thus also
the total) energy in the affected region is conserved.
Of course, this algorithm cannot accout for all the above mentioned
effects associated with multi-dimensional fluid motions, but at least one
can test the sensitivity of the post-bounce supernova evolution to
such a manipulation.

As a consequence of the mixing and the thus increased postshock pressure,
the hydrodynamic shock --- which in the absence of mixing reached a 
maximum radius of about $350$~km before it started to recede again ---
is indeed pushed out to a radius beyond 600~km
(Fig.~\ref{rampp.fig3}b), presumably leading to a weak supernova explosion. 
We interpret this result as an interesting hint that successful neutrino-driven
supernova explosions might be in reach when state-of-the-art neutrino
physics and a Boltzmann treatment of the neutrino transport are
eventually combined in self-consistent multi-dimensional simulations.

\subsection*{Acknowledgements}

It is a pleasure to thank K.~Takahashi for implementing the
improved treatment of the neutrino-nucleon cross sections and 
Ch.~Horowitz for providing correction formulae for the weak magnetism.
We thank the Institute for Nuclear Theory at the University 
of Washington for its hospitality and the Department of Energy
for support during a visit of the Summer Program on Neutron Stars.
This work was also supported by the Sonderforschungsbereich~375 on
``Astroparticle Physics'' of the Deutsche Forschungsgemeinschaft.
The computations were performed on the NEC SX-5/3C of
the Rechenzentrum Garching and on the CRAY T90 of the John 
von Neumann Institute for Computing (NIC) in J\"ulich.



\end{document}